# Rate and Aging Time Dependent Static Friction of a Soft and Hard Solid Interface

Vinay A. juvekar[1] and Arun K. Singh[2]
[1]Department of Chemical Engineering, IIT Bombay, Mumbai, India
[2]Department of Mechanical Engineering, VNIT Nagpur, India
E-mail:vaj@iitb.ac.in, aksinghb@gmail.com

## Abstract

In this article, we present a mathematical model that answers a classical question concerning how much force, which is generally called static friction force, will it require to initiate the motion of a soft solid block such as gel, rubber or elastomer on a hard surface for instance glass surface. The model uses population balance of the bonds between the polymer chains of the soft solid and the hard surface to estimate rate and aging time dependent static friction. The model predicts that under certain range of the pulling velocity, the friction stress at the onset of sliding (static friction stress) varies as the logarithm of the pulling velocity, as well as the logarithm of the aging time. These predictions are consistent with the experimental observations.

## 1. Introduction:

Friction of a soft solid such as a gel, rubber or elastomer, sliding on a hard surface, for example glass surface, arises due to bonding and debonding of molecular chains at the contacting interface.[1-11] This fact has been incorporated in the Schallamach friction model.[5] The bonds between the polymer chains and the hard surface undergo thermally activated rupture or adherence.[5,9,10] The rate of rupture of these bonds is enhanced on application of external stress thus the process can be modeled using the Eyring rate equation.[9] The Schallamach model has been widely used in explaining friction and adhesion phenomena at soft surfaces. For example, rate dependent increase of interfacial fracture energy[3,9] and steady dynamic friction stress.[1,7,9,11] Despite all these studies in the literature, we do not find in literature a static friction model derived from the first principle to answer a classical question that how much force will be required to initiate the onset of motion of a soft solid on a hard surface. In this paper, we have proposed a static friction model using population balance equation[1,6,11] (PBE) for bonds between the large population of polymer chains and the hard surface.

## 2. Modeling of static friction:

We consider the case where a soft solid block is allowed to age on a smooth hard surface for a time interval $t_w$. This time interval is generally called the aging time or waiting time or time of stationary contact or even relaxation time.[1,7] It is then that the upper face of the soft block is pulled with a constant velocity $V_0$, in the direction parallel to the base. During the waiting time, polymer chains adsorb on the hard surface. The number of bonds, between the polymer chains and the hard surface, increases with the aging or waiting time. During the pulling stage, the block undergoes shear deformation, as a result shear stress causes stretching of the chains which are bonded to the hard surface. The force generated due to stretching of the chains exactly balances the pulling force and this stress also causes creep deformation. The creep velocity $V_c$ progressively increases with time due to the increase in shear stress as well as decrease in the number of the live bonds due to breakage. As long as $V_c < V_0$, the block continues to deform and the stress in the soft block increases with time and the peak stress is reached when $V_c = V_0$. At this point, the stress is momentarily independent of time. We call this point as the point of onset of sliding though the actual sliding of the block has begun much earlier. The peak stress is the stress of static friction or threshold friction. Beyond the peak, $V_c > V_0$, and both the deformation of the block and the consequent stress decreases either to stable (dynamic friction) or unstable(stick-slip) motion.

During the waiting period, bonds are only formed. They neither age nor do they break if there is no external force or deformation on the soft block. Hence all bonds have zero age and the population balance equation[1,6,11] (PBE) simplifies to

$$\frac{dN(t)}{dt} = \frac{1}{\tau}\left[N_0 - N(t)\right] \tag{1}$$

Here $N(t)$ is the total number of live bonds and $N_0$ is the total number of available bonding sites. This equation yields the following solution subject to the initial condition, $N(t) = 0$ at $t = 0$.

$$N(t) = N_0\left[1 - exp\left(-\frac{t}{\tau}\right)\right] \tag{2}$$

The number $N_i$ of bonds, formed at the end of the waiting time $t_w$ is

$$N_i = N_0\left[1 - exp\left(-\frac{t_w}{\tau}\right)\right] \tag{3}$$

During the pulling process, the block is subjected to shear deformation. We analyze the dynamics of bonds during the time interval from the beginning of pulling to the time when the friction stress reaches the peak. We denote this time interval by $t_p$. The case of practical importance is the one for which $t_p << t_w$. Presently, we only analyze this case.

We assume that only those bonds, which are formed during the waiting period, bear the stress during the pulling period. We ignore the contributions from the bonds which are newly formed during the pulling period for three reasons. First, since the pulling time is much shorter than the waiting time, the number of bonds formed during the pulling time is much smaller. Second, the average age of the newly formed bonds is also small and hence they can bear lesser amount of load. Third, during the pulling process, the base of the block is continuously sliding hence time allowed for the bond formation is very short. The resulting bonds formed are, therefore, weaker than those formed during the waiting time when the block was stationary.

Since we ignore the bonds which are formed during the pulling period, we can drop the birth-term in the standard PBE model. Also, since all bonds formed during the aging period have the same age, we can write $t_a = t$ for all bonds and simplify the PBE to the following form

$$\frac{dN(t)}{dt} = -N(t)\frac{u}{\tau}e^{\lambda f(t)/kT} \tag{4}$$

$N(t)$ now represents the total number of live bonds at any time $t$ during the pulling phase. Negative sign on the right hand side of Eq.4 shows that $N(t)$ continuously decreases with time. We assume validity of the Hooke's law and write as

$$\frac{df(t)}{dt} = MV_c(t) \tag{5}$$

The total force is given by

$$F(t) = N(t)f(t) \tag{6}$$

The force $F(t)$ also equals the force generated in the block due to its shear deformation, and is obtained by modification of Eq.6 in the following form

$$\frac{dF(t)}{dt} = K_g \frac{d\Delta L(t)}{dt} = K_g (V_0 - V_c(t)) \tag{7}$$

Differentiating Eq.6 with respect to time and simplifying the resulting equation using Eqs.7 and 4, we obtain Eq.8 as

$$\frac{dF(t)}{dt} = -N(t)\frac{u}{\tau} f(t) e^{\lambda f(t)/kT} + N(t) M V_c(t) \tag{8}$$

After Eliminating $dF(t)/dt$ between Eqs. 7 and 8, we reach the following expression for the creep velocity

$$V_c(t) = \frac{K_g V_0 + N(t)\left(\frac{u}{\tau}\right) f(t) e^{\lambda f(t)/kT}}{N(t)M + K_g} \tag{9}$$

Substitution of this expression into Eq.7 yields

$$\frac{dF(t)}{dt} = K_g N(t) \left( \frac{M V_0 - \left(\frac{u}{\tau}\right) f(t) e^{\lambda f(t)/kT}}{N(t)M + K_g} \right) \tag{10}$$

We now eliminate $f(t)$ from Eqs.8 and 9 in the light of Eq.10 and modify them respectively to the following forms

$$\frac{dN(t)}{dt} = -N(t)\left(\frac{u}{\tau}\right) exp\left(\frac{\lambda F(t)}{N(t)kT}\right) \tag{11}$$

$$\frac{dF(t)}{dt} = K_g N(t) \left( \frac{M V_0 - \frac{u}{\tau N(t)} F(t) exp\left(\frac{\lambda F(t)}{N(t)kT}\right)}{N(t)M + K_g} \right) \tag{12}$$

The ordinary differential equations in Eqs.11 and 12 now contain only two dependent variables $F(t)$ and $N(t)$. They can be simultaneously solved using the following initial conditions

$$N(0) = N_i \quad and \quad F(0) = 0 \tag{13}$$

Where, $N_i$ is the number of bonds formed at the end of the aging period which is given by Eq. 3. We now covert Eqs.11 and 12 into the dimensionless form using the following dimensionless parameters

$$\hat{\sigma} = \frac{\sigma}{\sigma^*}, \hat{V} = \frac{V}{V^*}, \quad \hat{N} = \frac{N}{N_0}, \quad X = \frac{N_i - N}{N_i} \text{ and } r_s = \frac{N_0 M}{K_g} \tag{14}$$

In the above equation, $X$ is the fraction of the initial bonds which are broken at time $t$ and $r_s$ is the ratio of the total stiffness of the chains to the stiffness of the bulk material. Using these transformations, Eqs. 11 and 12 can now be modified to the following forms

$$\frac{dX}{d\hat{t}} = r_s u(1 - X) exp\left(\frac{\hat{\sigma}(\hat{t})}{(1 - X)\hat{N}_i}\right) \tag{15}$$

$$\frac{d\hat{\sigma}}{d\hat{t}} = \frac{\hat{V}_0 \hat{N}_i (1 - X) - \hat{\sigma} u \, exp\left(\frac{\hat{\sigma}(\hat{t})}{(1 - X)\hat{N}_i}\right)}{\hat{N}_i (1 - X) + 1/r_s} \tag{16}$$

The initial conditions given by Eq.13 transform to

$$X(0) = 0 \; and \quad \hat{\sigma}(0) = 0 \tag{17}$$

The velocity of creep can be written in dimensionless form as

$$\hat{V}_c(t) = \frac{\hat{V}_0 + u r_s \hat{\sigma}(t) exp\left(\frac{\hat{\sigma}(\hat{t})}{(1 - X)\hat{N}_i}\right)}{r_s \hat{N}_i (1 - X) + 1} \tag{18}$$

Further Eqs. 15 and 16 may also be written as

$$\frac{dX}{d(u\hat{t})} = r_s(1-X)exp\left(\frac{\hat{\sigma}(\hat{t})}{(1-X)\hat{N}_i}\right) \tag{19}$$

$$\frac{d\hat{\sigma}}{d(u\hat{t})} = \frac{\left(\frac{\hat{V}_0}{u}\right)\hat{N}_i(1-X) - \hat{\sigma}\, exp\left(\frac{\hat{\sigma}(\hat{t})}{(1-X)\hat{N}_i}\right)}{\hat{N}_i(1-X)+(1/r_s)} \tag{20}$$

Since $u$ is dimensionless, $u\hat{t}$ and $\hat{V}_0/u$ are also dimensionless and can be viewed as scaled time and scaled velocity, respectively. The advantage of using these scaled parameters is that the resulting equations do not explicitly contain $u$ and so the solutions expressing $\hat{\sigma}$ and $X$ in terms of $u\hat{t}$ and $\hat{V}_0/u$ are valid for all values of $u$.

**3. Results and discussion**: We have solved the dimensionless coupled differential equation (Eqs.19 and 20) numerically using the MATLAB solver *ode45* and the results are presented in the following Figs.1-5. Fig.1 shows typical plots of $\hat{\sigma}$, $1-X$ and $\hat{V}_c/u$, versus the scaled time $u\hat{t}$.

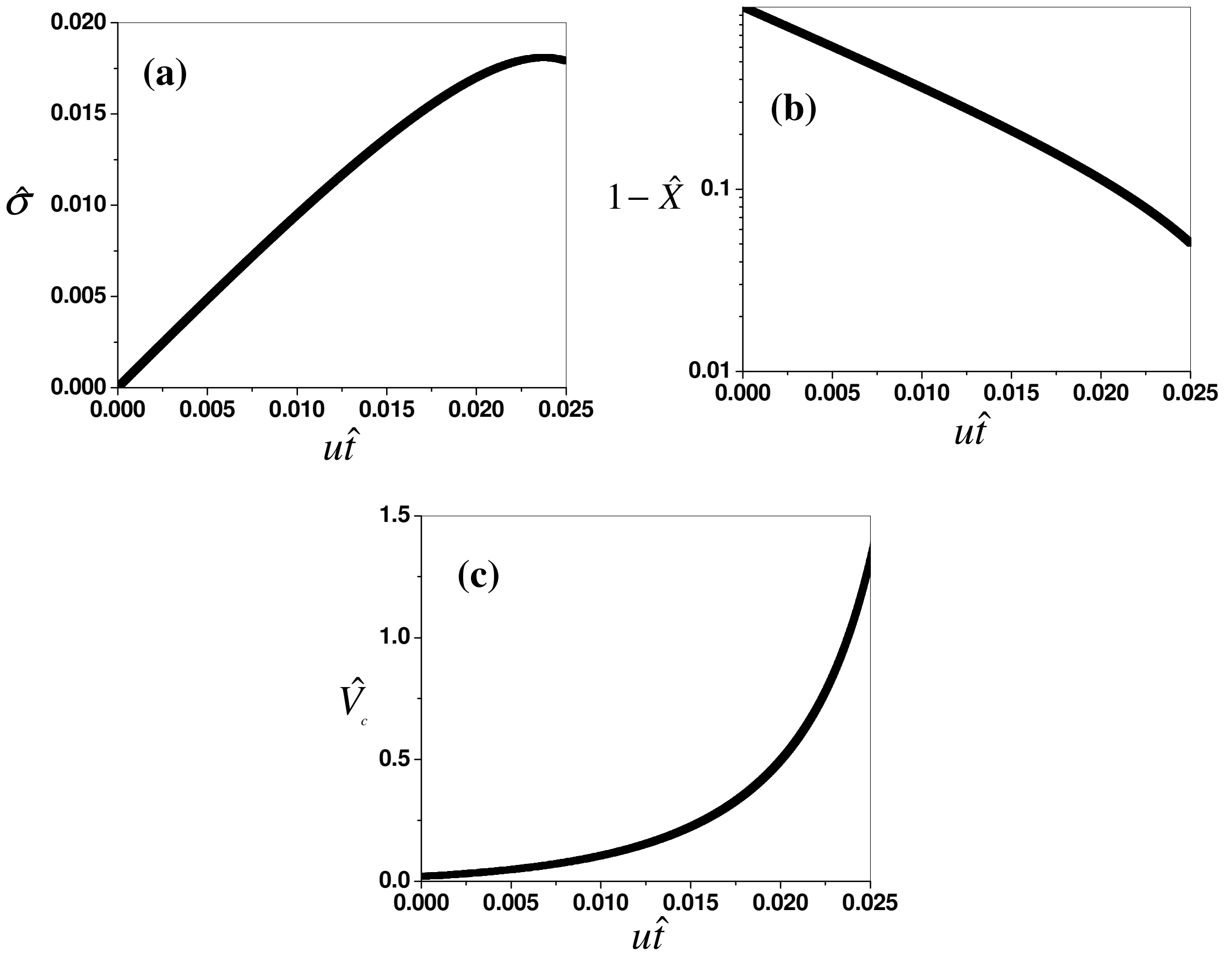

Fig.1. Pulling of a static block (a) Dimensionless stress, $\hat{\sigma}$ (b) Fraction of live bonds $1-X$ (c) Scaled dimensionless velocity of creep, $\hat{V}_c/u$, Each is plotted versus dimensionless scale time, $u\hat{t}$. Parameters: $\hat{V}_0/u=1$, $r_s=100$ and $\hat{N}_i=0.5$.

The initial creep velocity is obtained by substituting $X(0)=0$ and $\hat{\sigma}(0)=0$ in Eq. 18

$$\hat{V}_c(0)=\frac{\hat{V}_0}{r_s\hat{N}_i+1} \tag{21}$$

In general $r_s>>1$, $\hat{V}_c(0)<<\hat{V}_0$. This means that, in most practical situations, the initial velocity of creep is very small and the specimen undergoes only shear deformation. This is clear from Fig. 1(c). As the point of static friction is approached, there is a rapid rise in $\hat{V}_c$. Moreover, during the initial period, the magnitude of the stress is small, consequently Eq.19 can be simplified by dropping the second term in the numerator and $1/r_s$ in the denominator to yield

$$\frac{d\hat{\sigma}}{d(u\hat{t})}=\left(\frac{\hat{V}_0}{u}\right) \tag{22}$$

The plot of $\hat{\sigma}$ versus the scaled time $u\hat{t}$ should be a straight line with the magnitude of slope equal to $\hat{V}_0/u$. Fig.1 (a) confirms that the initial portion of the plot is indeed a straight line with slope equal to 1, which is basically the value of $\hat{V}_0/u$ used.

We also conclude from Eq.19 that the initial rate of bond breakage can be approximated as

$$\frac{dX}{d(u\hat{t})}=r_s(1-X) \tag{23}$$

The plot of $ln(1-X)$ versus the scaled time $u\hat{t}$ should be a straight line with slope $-r_s$. This is also evident from Fig.1(b) which plots $1-X$ on logarithmic scale versus $u\hat{t}$ on linear scale.

In Fig.2, we have plotted the stress $\hat{\sigma}$ as a function of $u\hat{t}$ for different pulling velocity $\hat{V}_0$ on a Log-Log scale. The plots are straight lines with unit slope almost up to the peak point, indicating the validity of Eq.22. It is also clear from the plots in Fig.2 that as the $\hat{V}_0$ increases, the static friction (peak) stress $\hat{\sigma}_P$ increases. On the other hand, the pulling time $\hat{t}_p$ decreases as evident in Fig.2. The reason for this behavior is that at higher $\hat{V}_0$, the pulling stress increases more rapidly thereby a given value of the stress is attained in shorter time. Consequently, the bonds are allowed shorter time to break compared to the case where pulling velocity is lower. Fewer bonds are therefore broken and more bonds are available to resist the force, yielding higher peak stress.

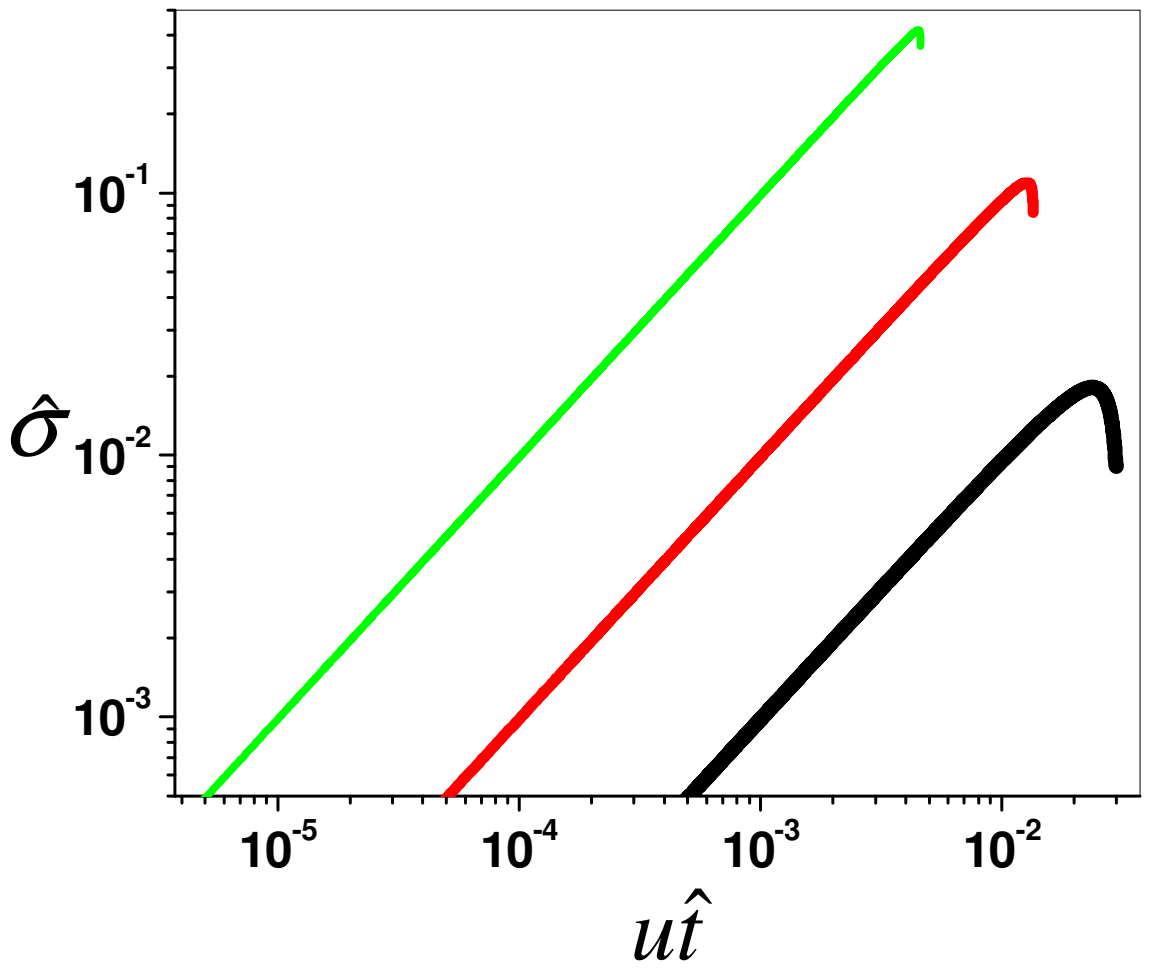

Fig.2. Plot of dimensionless stress $\hat{\sigma}$ versus scaled time $u\hat{t}$ during static friction experiment, effect of the pulling velocity. $\hat{N}_i = 0.5$, $r_s = 100$, $\hat{V}_0/u = 1$ (black), $\hat{V}_0/u = 10$ (red), $\hat{V}_0/u = 100$ (green).

Fig.3 plots the stress $\hat{\sigma}$ versus $u\hat{t}$ using the initial extent of bonding $(\hat{N}_i)$ as the parameter. It is seen from the plots in Fig.3 that as $\hat{N}_i$ decreases, the static stress decreases and the peak time $\hat{t}_p$ also decreases and this is expected.

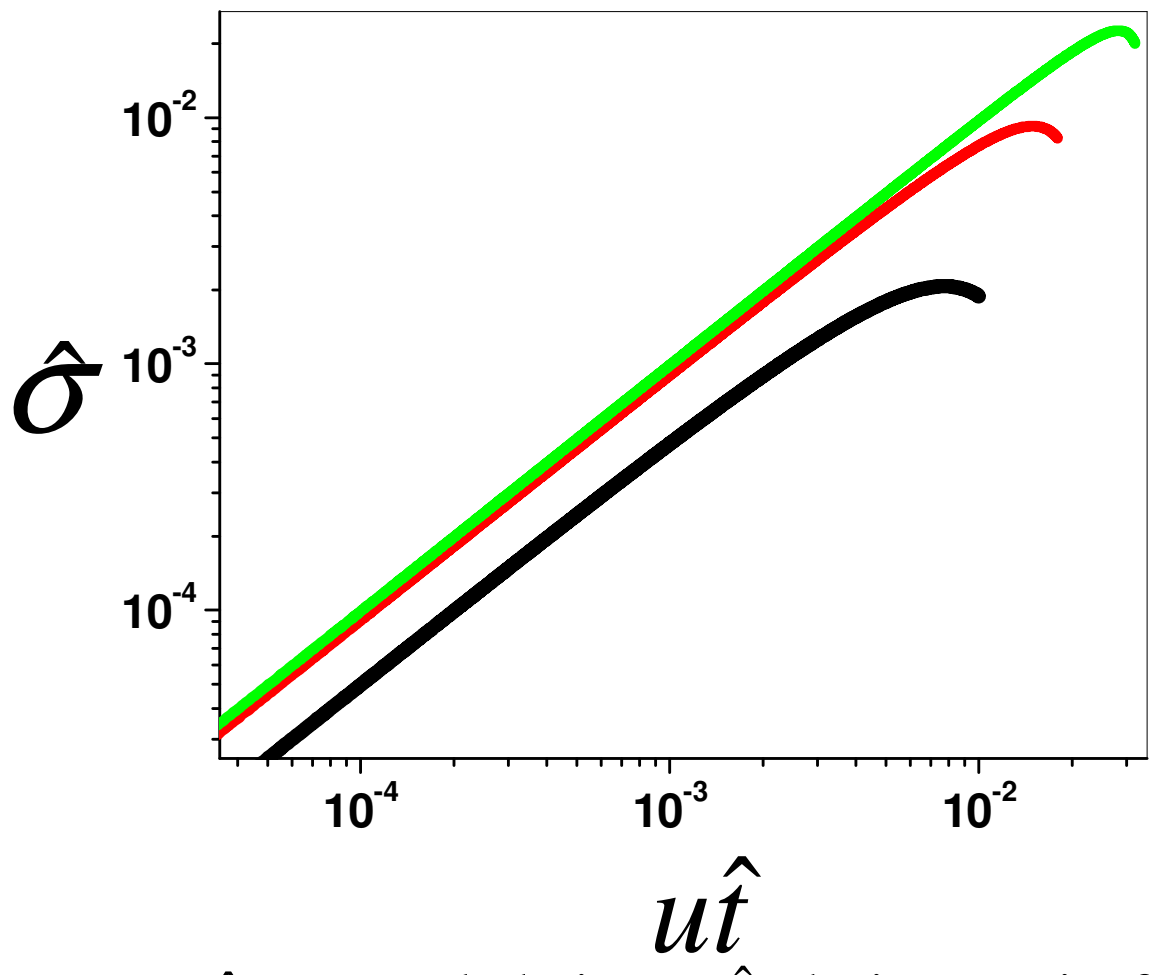

Fig.3. Dimensionless stress $\hat{\sigma}$ vs. scaled time $u\hat{t}$ during static friction experiment, effect of $\hat{N}_i$ for $\hat{N}_i = 0.01$ (black) $\hat{N}_i = 0.1$ (red), $\hat{N}_i = 1$ (green), $\hat{V}_0 = 1, r_s = 100$.

We also see that the initial stress decreases with decreasing $\hat{N}_i$, resulting in a horizontal shift of the curves. The reason is, when lesser number of bonds is initially formed, the initial creep velocity is higher by Eq.21 and difference between the velocity of the puller and the base of the specimen is lower, thus lowering the shear stress. The effect is pronounced for $\hat{N}_i = 0.01$, since in this case $r_s\hat{N}_i$ in Eq.21 is comparable to unity. More significantly, since the static friction stress $\hat{\sigma}_P$ represents the maximum stress and $d\hat{\sigma}/d\hat{t} = 0$ at that point. From Eq.22, we conclude that peak stress satisfies the following equation

$$\hat{\sigma}_p e^{\frac{\hat{\sigma}_p}{(1-X_p)\hat{N}_i}} = \left(\frac{\hat{V}_0}{u}\right)\hat{N}_i(1-X_p) \quad (24)$$

The equation has two unknowns and hence cannot be solved in isolation. We can, however, obtain an approximate solution of Eq.24 by assuming the approximation of Eqs.21 and 22 to be valid up to the peak stress $\hat{\sigma}_P$ and combine them to get

$$1 - X_p = exp\left(-\frac{r_s u \hat{\sigma}_p}{\hat{V}_0}\right) \quad (25)$$

Elimination of $X_p$ between Eq.24 and Eq.25 results in the following equation

$$\frac{\hat{\sigma}_p}{\hat{N}_i} e^{\frac{r_s u \hat{\sigma}_p}{\hat{V}_0}} + \frac{r_s u \hat{\sigma}_p}{\hat{V}_0} + ln(\hat{\sigma}_p) = ln\left(\frac{\hat{V}_0 \hat{N}_i}{u}\right) \tag{26}$$

This is an approximate relation but a direct relation between the friction stress and the pulling velocity.

The values of the static friction stress $\sigma_p$, obtained from the solution of Eqs.25 and 26, are plotted against the scaled pulling velocity $\hat{V}_0/u$ in Fig.4. They are shown by solid lines at low and high ranges of velocities. They are compared with the approximate solution obtained by solving Eq.26 and found to be in good agreement.

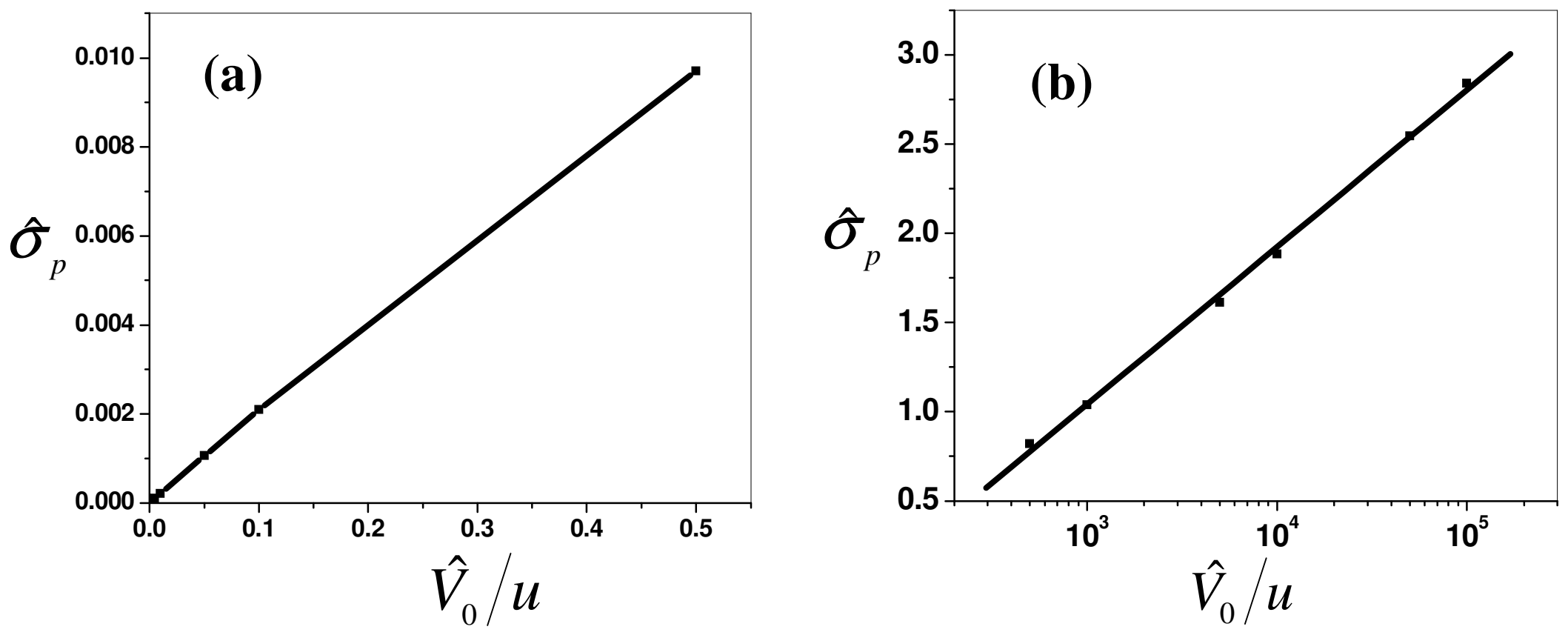

Fig.4. Plot of static friction force $\hat{\sigma}_P$ versus the scaled pulling velocity $\hat{V}_0/u$ (a) Low velocity range. (b) High velocity range for friction parameters: $\hat{N}_i = 0.5, r_s = 100$.

Two distinct trends are observed in parts (a) and (b) of Fig.4. At low velocities, the static friction stress varies linearly with $\hat{V}_0/u$ and at high velocities, it varies logarithmically with $(\hat{V}_0/u)$. We justify these trends using the approximate Eq.26. At very low pulling velocity, the first (exponential) term in Eq.26 dominates. The exponent $r_s u \hat{\sigma}_p/\hat{V}_0$ should therefore change very slowly in order that the equation is satisfied. The constancy of the exponent implies linear variation of $\hat{\sigma}_p$ with $(\hat{V}_0/u)$. On the other hand, for large $(\hat{V}_0/u)$, the exponent tends to zero and Eq.26 reduces to

$$\frac{\hat{\sigma}_P}{\hat{N}_i} = ln\left(\frac{\hat{V}_0 \hat{N}_i}{u}\right) \quad (27)$$

which is a logarithmic relation. The logarithmic dependence of the static stress on pulling velocity has been observed experimentally[1,3,10].

The effect of waiting or aging time on $\hat{\sigma}_P$ can be obtained by noting that the initial number of bonds $\hat{N}_i$ is related to the waiting or aging time by Eq.3, which can be written in dimensionless form as

$$\hat{N}_i = \left[1 - exp\left(- r_s \hat{t}_w\right)\right] \quad (28)$$

Fig.5 plots the static friction $\hat{\sigma}_P$ versus the normalized waiting time, $\hat{t}_w$. The waiting time is plotted on logarithmic scale. It is seen that the plot in Fig.5 has a sigmoid shape. At very long waiting time, the value of $\hat{N}_i \to 1$ and hence the static friction force reaches a plateau. At very short waiting time, $\hat{N}_i \to 0$ and the static friction force tends to zero. There is a point of inflection.

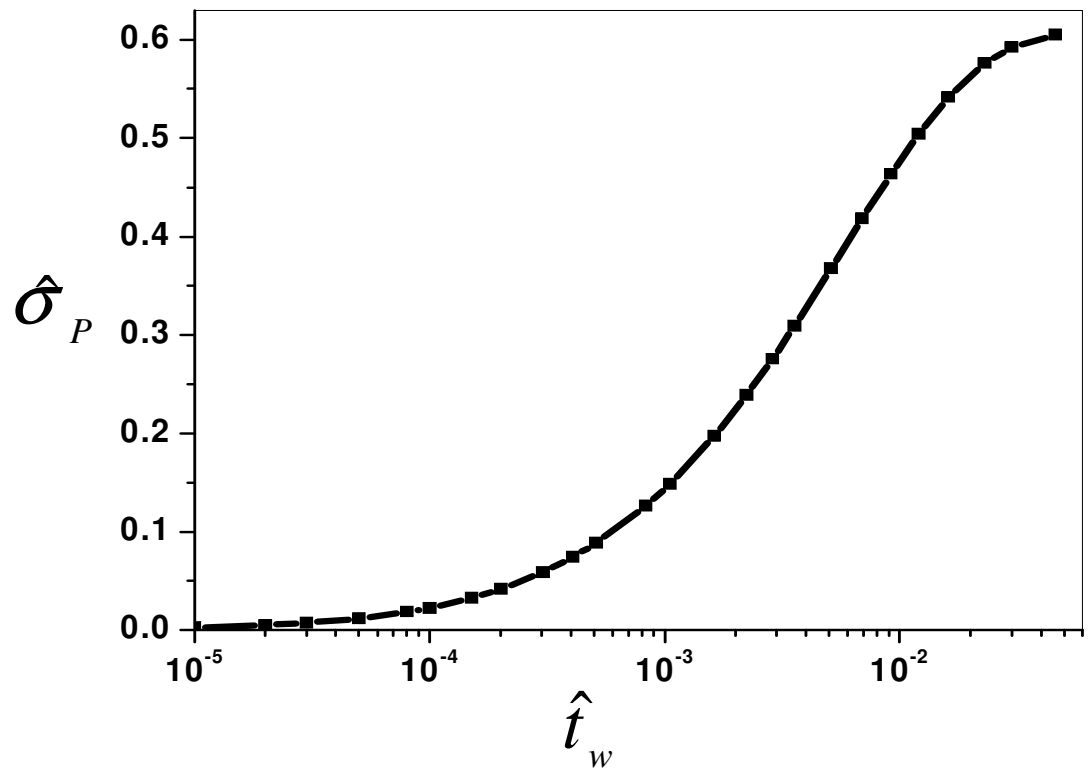

Fig.5. Plot of static friction stress versus the scaled waiting time for $\hat{V}_0/u = 100, r_s = 100$.

A large portion of the curve in the vicinity of the point of inflection can be approximated by a straight line which is basically logarithmic behavior of static friction. This portion of the plot corresponds to approximately $0.2 < \hat{N}_i < 0.8$. The logarithmic dependence of the static tress on waiting time has been observed experimentally[1].

Although present technique of modeling static friction has been used for a soft and hard solid interface, the procedure is general in nature and can be used for other sliding surfaces for instance hard-hard solid interface at which large number of micro contacts give rise to friction[12].

**4. Conclusion:**

In the present work, we have developed a population balance based model for estimating static friction of a soft solid on a hard surface. It is shown theoretically that static friction stress varies as logarithm of pulling velocity as well as logarithm of waiting time or time of stationary contact. These results are consistent with the experimental observations.